\documentclass[reprint, amsmath, amssymb, superscriptaddress, pra]{revtex4-2}
\usepackage{graphicx}
\usepackage{hyperref}
\usepackage{xcolor}
\usepackage{bbm}
\usepackage{stmaryrd}

\begin{document}
\title{Tuning and Amplifying the Interactions in Superconducting Quantum Circuits \\
with Subradiant Qubits}

\author{Qi-Ming Chen}
\email{qiming.chen@wmi.badw.de}
\affiliation{Walther-Mei{\ss}ner-Institut, Bayerische Akademie der Wissenschaften, 85748 Garching, Germany}
\affiliation{Physik-Department, Technische Universit{\"a}t M{\"u}nchen, 85748 Garching, Germany}

\author{Florian Fesquet}
\author{Kedar E. Honasoge}
\author{Fabian Kronowetter}
\author{Yuki Nojiri}
\author{Michael Renger}
\author{Kirill G. Fedorov}
\affiliation{Walther-Mei{\ss}ner-Institut, Bayerische Akademie der Wissenschaften, 85748 Garching, Germany}
\affiliation{Physik-Department, Technische Universit{\"a}t M{\"u}nchen, 85748 Garching, Germany}

\author{Achim Marx}
\affiliation{Walther-Mei{\ss}ner-Institut, Bayerische Akademie der Wissenschaften, 85748 Garching, Germany}

\author{Frank Deppe}
\email{frank.deppe@wmi.badw.de}
\author{Rudolf Gross}
\email{rudolf.gross@wmi.badw.de}
\affiliation{Walther-Mei{\ss}ner-Institut, Bayerische Akademie der Wissenschaften, 85748 Garching, Germany}
\affiliation{Physik-Department, Technische Universit{\"a}t M{\"u}nchen, 85748 Garching, Germany}
\affiliation{Munich Center for Quantum Science and Technology (MCQST), Schellingstr. 4, 80799 Munich, Germany}

\date{\today}
\begin{abstract}
We propose a tunable coupler consisting of $N$ fixed-frequency qubits, which can tune and even amplify the effective interaction between two superconducting quantum circuits. The tuning range of the interaction is proportional to $N$, with a minimum value of \textit{zero} and a maximum that can exceed the physical coupling rates between the coupler and the circuits. The effective coupling rate is determined by the collective magnetic quantum number of the qubit ensemble, which takes only discrete values and is free from collective decay and decoherence. Using single-photon $\pi$-pulses, the coupling rate can be switched between arbitrary choices of the initial and final values within the dynamic range in a single step without going through intermediate values. A cascade of the couplers for amplifying small interactions or weak signals is also discussed. These results should not only stimulate interest in exploring the collective effects in quantum information processing, but also enable development of applications in tuning and amplifying the interactions in a general cavity-QED system.
\end{abstract}

\maketitle

\section{Introduction}
Tuning the coupling rate between two fixed-frequency superconducting quantum circuits, instead of tuning the characteristic frequency of each individual part, has attracted an increasing interest in recent years as a promising way to scalable quantum computing \cite{Niskanen2007, Chow2011, McKay2016, Caldwell2018, Arute2019, Foxen2020, Xu2020, Bengtsson2020, Collodo2020}. The conventional method to realize such a tunable coupler is to place a Josephson junction, or SQUID, as a mediating element between the two circuits. From a circuit point of view, the junction can be regarded as a tunable positive inductance which, together with other circuit elements such as capacitors or mutual inductors, can be used to form a tunable coupler \cite{Blais2003, Gambetta2011, Bialczak2011, Chen2014, Geller2015}. Alternatively, one may also consider the junction as an off-resonant qubit which mediates the exchange of virtual photons between the two circuits \cite{Sun2006, Mariantoni2008, Srinivasan2011, Baust2015, Yan2018, Mundada2019, Li2020, Han2020}. A tunable coupling rate is achieved by adjusting the frequency detuning between the coupler qubit and the coupled circuits, while maintaining the former unexcited. This single-junction coupler has attracted a huge success in recent experiments. However, the dynamic range of the effective component-component coupling rate is limited to the second order of the physical dispersive coupling rate. Moreover, the coupler can be very sensitive to experimental imperfections, including both systematic and stochastic errors in certain parameter regimes, because of the nonlinearity of the Josephson inductance. 

Here, we propose to use different steady states of one or several Josephson junctions to tune the coupling rate between two superconducting quantum circuits. The coupler is modeled by an ensemble of $N$ homogeneous fixed-frequency and off-resonant qubits, also known as the Dicke model \cite{Dicke1954}, and is thus called the D-coupler. Instead of tuning the frequencies of the coupler qubits, a tunable interaction is achieved by preparing the qubits in different quantum states, corresponding to different collective angular and magnetic quantum numbers \cite{Mariantoni2008, Baust2015, Chen2018}. The dynamic range is proportional to the number of coupler qubits, where the maximum coupling rate may even exceed the individual coupling rates in the system for large $N$. On the other hand, the name, D-coupler, may also be interpreted as ``decay- and decoherence-free" if we prepare the qubits in subradiant states \cite{Dicke1954}. It can also be understood as a digital coupler if we further restrict the subradiant states to be pairwise. In this case, the effective coupling rate takes only discrete values that are proportional to the total excitation number of the qubit ensemble \cite{Chen2018}. Control of the coupling rate, between arbitrary initial and final choices within the dynamic range, can be realized in a single step with single-photon $\pi$-pulses \cite{Scully2015}. In the meanwhile, the coupling rate needs not to go through any intermediate values during the tuning process. These properties make the D-coupler an ideal device in superconducting quantum circuits, despite the potential technical difficulty in sample design and fabrication, and should motivate more applications of the collective effects in quantum information processing \cite{Fink2009, Filipp2011, Filipp2011a, Delanty2011, Loo2013, Mlynek2014, Ma2019, Song2019, Ye2019, Yan2019, Wang2020, Gong2021}.

\begin{figure*}[hbt]
  \centering
  \includegraphics[width=2\columnwidth]{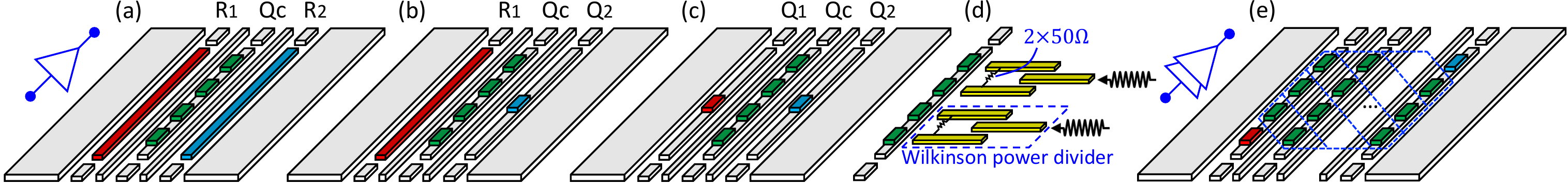}
  \caption{(a)-(c), Schematic of the D-coupler (green) between two resonators, $\rm R_1$ and $\rm R_2$, one resonator and one qubit, $\rm R_1$ and $\rm Q_2$, as well as two qubits, $\rm Q_1$ and $\rm Q_2$, which are colored in red and blue, respectively. The coupler is labelled as $\rm Q_{c}$, which is an ensemble of $N$ qubits. The effective coupling rate between the two circuits is determined by the collective magnetic quantum number, $m$, of $\rm Q_{c}$. (d), A change of the magnetic quantum number, $m \rightarrow m'$, can be realized in a single step by applying $(m'-m)$ single photons to $(m'-m)$ Wilkinson power dividers, each of which couples to two qubits in the ensemble with a $\pi$ phase difference. We note that the qubit number in the ensemble, $N$, is assumed to be \textit{even}, while an \textit{odd} $N$ leads to a finite minimum coupling rate corresponding to the magnetic quantum numbers $\pm 1/2$. (e), A cascade of $D$ layers of D-couplers (green) to mediate the interaction between two qubits colored in red and blue.}
  \label{fig:schematic}
\end{figure*}

\section{The Dicke coupler}
We consider a system where the interaction between two circuits, $\rm X_1$ and $\rm X_2$, is mediated by an ensemble of $N$ qubits, $\rm Q_{c}$. The system Hamiltonian may be written as ($\hbar=1$) \cite{Blais2004}
\begin{align}
	H &= \sum_{m=1}^{2}\sum_{n=1}^{N}\omega_{m} x_m^{\dagger}x_m 
	+ \frac{\omega_{n}}{2}\sigma_{n}^{z}
	+ g_{m,n}\left(x_{m}^{\dagger} + x_{m}\right)\sigma_{n}^{x} \nonumber \\
	&+ \sum_{(n, n')} g_{n,n'} \left(\sigma_{n}^{+}\sigma_{n'}^{-} + \sigma_{n}^{-}\sigma_{n'}^{+}\right).
	\label{eq:hamiltonian}
\end{align}
Here, $\omega_{m}$, $x_m$, and $x_m^{\dagger}$ are the resonant frequency, raising, and lowering operators of circuit $\rm X_m$. Furthermore, $\omega_{n}$ and $\sigma_{n}^{\alpha}$ with $\alpha=x,y,z$ are the characteristic frequency and the standard Pauli operators of the $n$th coupler qubit, $g_{\alpha,\beta}$ is the physical coupling rate between the two parts $\alpha$ and $\beta$, and $(n, n')$ takes all possible pairs in the ensemble. Depending on the commutation and anti-commutation relations between $x_m$ and $x_m^{\dagger}$, the above model can be used to describe a resonator-resonator (R-R) coupler ($x_1 \equiv r_1$, $x_2 \equiv r_2$), a resonator-qubit (R-Q) coupler ($x_1 \equiv r_1$, $x_2 \equiv \sigma_{2}^{-}$), or a qubit-qubit (Q-Q) coupler ($x_1 \equiv \sigma_{1}^{-}$, $x_2 \equiv \sigma_{2}^{-}$), as schematically shown in Fig.\,\ref{fig:schematic}(a)-(c).

Assuming that the coupler qubits are largely detuned from the two circuits to be coupled, i.e., $g_{m,n} \ll \Delta_{m,n}, \Sigma_{m,n}$ where $\Delta_{m,n}=\omega_{m}-\omega_{n}$ and $\Sigma_{m,n}=\omega_{m}+\omega_{n}$, a dispersive approximation (DA) may be applied to Eq.\,\eqref{eq:hamiltonian} that diagonalizes the coupler degree of freedom to the second order of $g_{m,n}$. For simplicity and in the same spirit of the Dicke \cite{Dicke1954} or Tavis-Cummings model \cite{Tavis1968, Tavis1969}, we further assume the qubits to be homogeneous and use the collective angular momentum operators, $J^{\alpha} = \sum_{n=1}^{N} \sigma_{n}^{\alpha}$ with $\alpha=x,y,z$, to describe the whole ensemble \cite{Lopez2007}. Then, the effective Hamiltonian can be written in a more compact form (see Appendix\,\ref{app:one} for detailed derivation)
\begin{align}
	\tilde{H} &= \sum_{m=1}^{2} \omega_{m} x_m^{\dagger}x_m 
	+ \frac{1}{2} \left[ \omega_{\rm c} + \chi_{m}^{-} \left(x_m + x_m^{\dagger} \right)^2 \right] J^{z} \nonumber \\
	&- \sum_{m=1}^{2} \frac{g_1g_2}{2}\left(\frac{1}{\Delta_{m}}-\frac{1}{\Sigma_{m}}\right)J^{z}
	\left( x_1^{\dagger} + x_1 \right)\left( x_2^{\dagger} + x_2 \right) \nonumber \\
	&+ \frac{g_{\rm c}}{2}\left(J^{+}J^{-}+J^{-}J^{+}\right)
	+ \frac{\chi_{m}^{+}}{2}\llbracket x_m, x_m^{\dagger}\rrbracket\left(J^{x}\right)^{2}, \label{eq:effective}
\end{align}
where $\chi_{m}^{\pm}=-g_m^2\left(1/\Delta_{m}\pm 1/\Sigma_{m}\right)$, $g_{m} \equiv g_{m,n}$, $g_{\rm c} \equiv g_{n,n'}$, and $\llbracket A,B \rrbracket=AB-BA$ is the commutation operator. One can verify that the angular quantum number, $j$, which is determined by $j(j+1)=\langle J^2 \rangle$, is a conserved quantity. However, the magnetic quantum number, $m$, which is an eigenvalue of $J^{z}/2$, may not be a good quantum number because of the counter-rotating term $\left(J^{x}\right)^2$. To be able to apply a rotating wave approximation (RWA), we further assume $\chi_{m}^{+} \ll \omega_{\rm q}$. Then, $J^{z}$ commutes with the Hamiltonian and can be replaced by its average during the time evolution if the system is initially prepared at its eigenstates. This results in an effective coupling rate between $\rm X_1$ and $\rm X_2$,
\begin{align}
	g_{\rm eff} = - \sum_{m=1}^{2} \frac{g_1g_2}{2}\left(\frac{1}{\Delta_{m}}-\frac{1}{\Sigma_{m}}\right)\langle J^{z} \rangle. \label{eq:effective_g}
\end{align}

The above result indicates that an arbitrary coupling rate, which corresponds to $-N \leq \langle J^{z}\rangle \leq N$, can be achieved by preparing the qubits in different collective states $|j,m\rangle$ with $j=0,\cdots,N/2$, $m=-j,\cdots,j$. One may also prepare the coupler in an arbitrary superposition or mixture of the eigenstates $|j,m\rangle$, which results into a quantum effective coupling strength between the two circuits and may leads to novel applications. However, we restrict the coupler to be eigenstates in the rest of this paper. For a sufficiently large $N$, it is possible to achieve an effective coupling rate exceeding the physical rates between the coupler and either of the circuits, $g_{\alpha,\beta}$, while maintaining the requirements of DA and RWA. These properties show that a qubit ensemble described by the Dicke model can be used as a tunable coupler between two circuits and has a broad dynamic range proportional to the size of the ensemble.

\begin{figure}[hbt]
  \centering
  \includegraphics[width=\columnwidth]{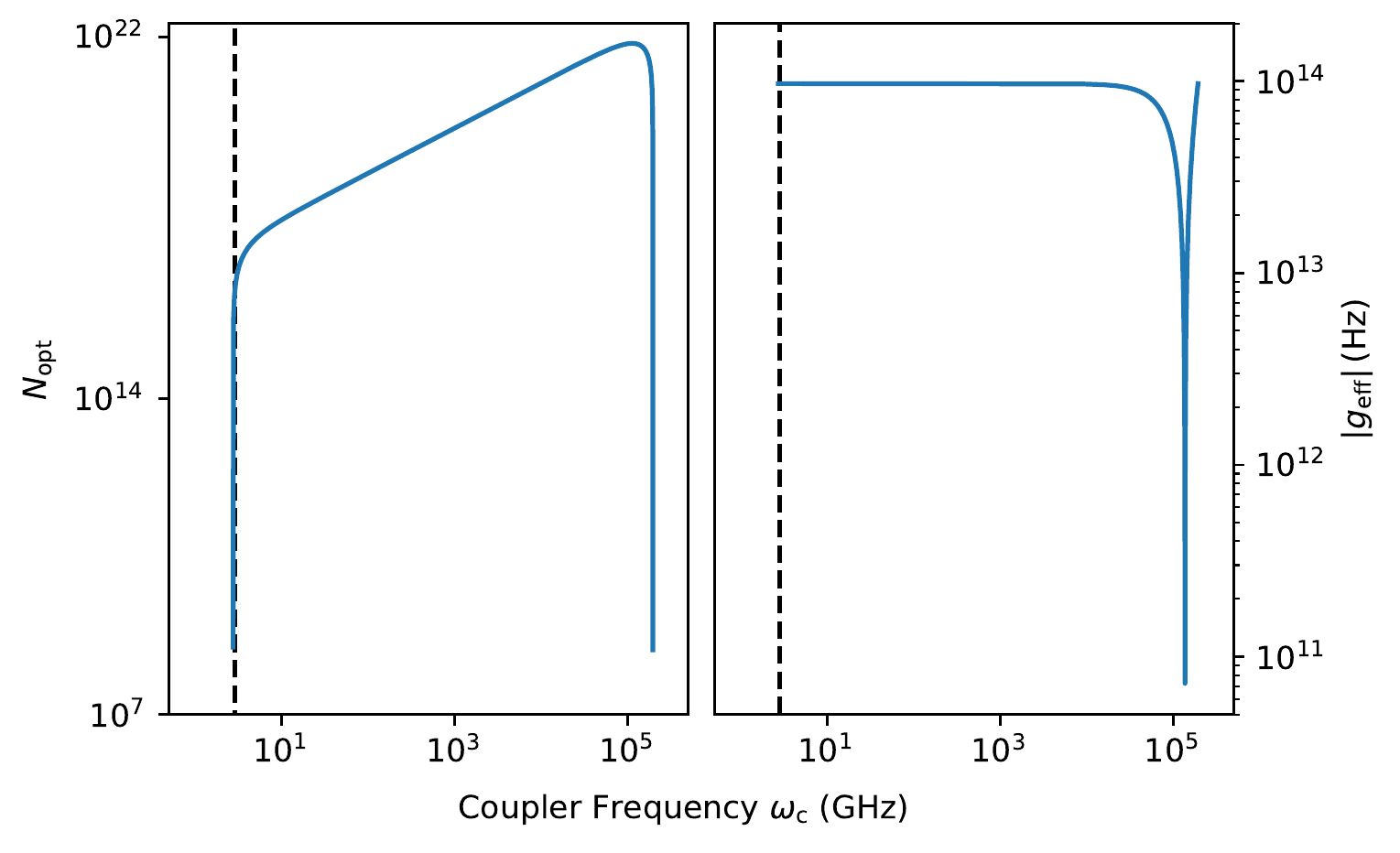}
  \caption{The optimal number of qubits, $N_{\rm opt}$, for microwave-optical photon conversion (left) and the absolute effective coupling rate, $\left|g_{\rm eff}\right|$, between the microwave and optical modes (right). Here, the coupler frequency, $\omega_{\rm c}$, is swept from the microwave to optical frequency. The dashed line indicates the frequency splitting between the spin-$\pm 1$ and the spin-$0$ states of an NV-center in the electron ground state.}
  \label{fig:transducer}
\end{figure}

We note that the frequencies of the two circuits being coupled are not necessarily the same, and the above discussions apply to a general system which can be described by the Dicke model. One may consider using the D-coupler as a transducer which converts a microwave photon to an optical photon, and \textit{vice versa} \cite{Sun2006}. Here, the qubit ensemble may be made of a thin layer of spin-$1/2$ materials, for example, a huge and homogeneous array of NV-centers, at an intermediate frequency between microwave and optical frequencies. At the optimal condition
\begin{align}
	\frac{\omega_{1}-\omega_2}{N} = g_{2}^2\left(\frac{1}{\Sigma_{2}} 
	- \frac{1}{\Delta_{2}}\right) - g_{1}^2\left(\frac{1}{\Sigma_{1}} 
	- \frac{1}{\Delta_{1}}\right),
\end{align}
where the two physically off-resonant modes are effectively on resonance such that a single photon can be perfectly transferred between them for a time duration of $t=\left(\pi+2k\pi\right)/\left[g_1g_2\left(1/\Delta_{1}-1/\Sigma_{1}+1/\Delta_{2}-1/\Sigma_{2}\right)N\right]$ for $k=0,1,\cdots$. This result may indicate a strong coupling rate between the microwave and optical modes for large $N$ and, correspondingly, a high conversion efficiency ideally approaching to the unity. 

For a qualitative estimation, we choose the frequency of the microwave and optical modes to be $2.744\,{\rm GHz}$ and $194.3\,{\rm THz}$, respectively \cite{Forsch2019}. The coupler frequency is $2.87\,{\rm GHz}$ for an ensemble of NV centers, with a coupling strength of $1\,{\rm kHz}$ between the coupler and either of the two modes. We estimate that the optimal condition leads to an effective coupling rate at the scale of $10^{14}\,{\rm Hz}$, which is in the deep strong coupling regime (Fig.\,\ref{fig:transducer} right). However, one should also note that the required number of homogeneous colored centers is approximately $2.4\times 10^{16}$, which makes the optimal condition an extreme challenge to reach in real experiments (Fig.\,\ref{fig:transducer} left). In these regards, a practical implementation of the transducer requires a compromise between the effective coupling rate and the optimal condition. Alternatively, one may consider using two harmonic oscillators to form an effective spin-$N/2$ coupler, which is equivalent to an ensemble of $N$ spin-$1/2$ qubits \cite{Schwinger1952, Sakurai1994}. In either of the two cases, the collective effect of the Dicke model plays an important role in amplifying the weak interaction between the two off-resonant modes. A more detailed discussion of such a transducer relies on the specific parameters of the system and the compromises one may take, and is thus beyond the major interest of this study. 

\section{The decay $\&$ decoherence-free coupler}
The D-coupler is free from collective decay and decoherence if one restricts the collective states to a subspace spanned by subradiant states \cite{Dicke1954}. By definition, a subradiant state is an eigenstate of the collective angular momentum, $J^{z}/2$, with eigenvalue $m=-j$. In this regard, it is also an eigenstate of the collective lowering operator, $J^{-}$, with eigenvalue \textit{zero}. In an open environment, the dynamics of the system may be described by the master equation with the Born-Markov approximation \cite{Carmichael1999}
\begin{align}
	\dot{\rho} = -i\llbracket H, \rho \rrbracket + \frac{\gamma}{2}\mathcal{D}\left[J^{-}\right]\rho + \frac{\gamma_{\phi}}{2}\mathcal{D}\left[J^{z}\right]\rho,
\end{align}
where $\gamma$ and $\gamma_{\phi}$ are the energy relaxation and dephasing rates of the qubit ensemble, respectively, and $\mathcal{D}\left[J^{\alpha}\right]\rho=2J^{\alpha}\rho \left(J^{\alpha}\right)^{\dagger} - \left(J^{\alpha}\right)^{\dagger}J^{\alpha}\rho - \rho\left(J^{\alpha}\right)^{\dagger}J^{\alpha}$ is the Lindblad superoperator. One can verify that the subradiant states are also eigenstates of the Lindblad superoperator with eigenvalue \textit{zero}. Thus, an arbitrary superposition of the subradiant states remains invariant during the dynamics of the open system with regard to collective decay and decoherence. Correspondingly, the effective coupling rate, which now takes only \textit{zero} or negative eigenvalues of $\langle J^{z} \rangle$, is potentially stable in an open environment. 

\section{The digital coupler}
Tuning of the coupling rate can be made ultrafast if we further restrict the subradiant states of the coupler to be pairwise, where each adjacent qubit pair $(2n, 2n+1)$ takes only ground or singlet states, denoted as $|g\rangle=|0_{2n}0_{2n+1}\rangle$ and $|s\rangle = \left(|1_{2n}0_{2n+1}\rangle - |0_{2n}1_{2n+1}\rangle\right)\sqrt{2}$ \cite{Scully2015}. However, the dynamic range of the effective coupling strength is not affected because the degeneracy of the subradiant states. This restriction also relaxes the original assumption of a collective decay and decoherence for all the coupler qubits into that for each qubit pair $(2n, 2n+1)$ \cite{Chen2018}. The coupling rate is controlled digitally by counting the total number of excitations in the qubit ensemble. Different from our earlier proposal for controlling the R-R coupling rate \cite{Chen2018}, we describe an alternative method that applies to a general system and results into an even faster switch of $g_{\rm eff}$ with beam splitters and single photon sources \cite{Scully2015}.

As schematically shown in Fig.\,\ref{fig:schematic}(d), the control protocol may consist of a Wilkinson power divider which is modeled as a beam splitter routing an incident single microwave photon to two different paths simultaneously but with a $\pi$-phase difference \cite{Mariantoni2010, Menzel2010, Menzel2012, Fedorov2016}. At the end of each path, we couple one qubit to it and describe the photon-qubit interaction by a Jaynes-Cummings model \cite{Jaynes1963}. Assuming that the two addressed qubits are initially in the ground state, $|g\rangle$, the single-photon drive increases the magnetic quantum number, $m$, by one and results in a singlet state $|s\rangle$ \cite{Scully2015}. Alternatively, if the qubits are initially in the singlet state, they will end up in the ground state with $m \rightarrow (m-1)$ \cite{Scully2015}. For an ensemble of $N/2$ qubit pairs, one can apply $N/2$ single-photon $\pi$-pulses to them in parallel, and switch the coupling from an arbitrary initial to the final value in a single step. Interestingly, the coupling rate needs not to go through any intermediate values during the tuning process if the skew among different beam splitters is negligibly small. 

\begin{figure*}
  \centering
  \includegraphics[width=2\columnwidth]{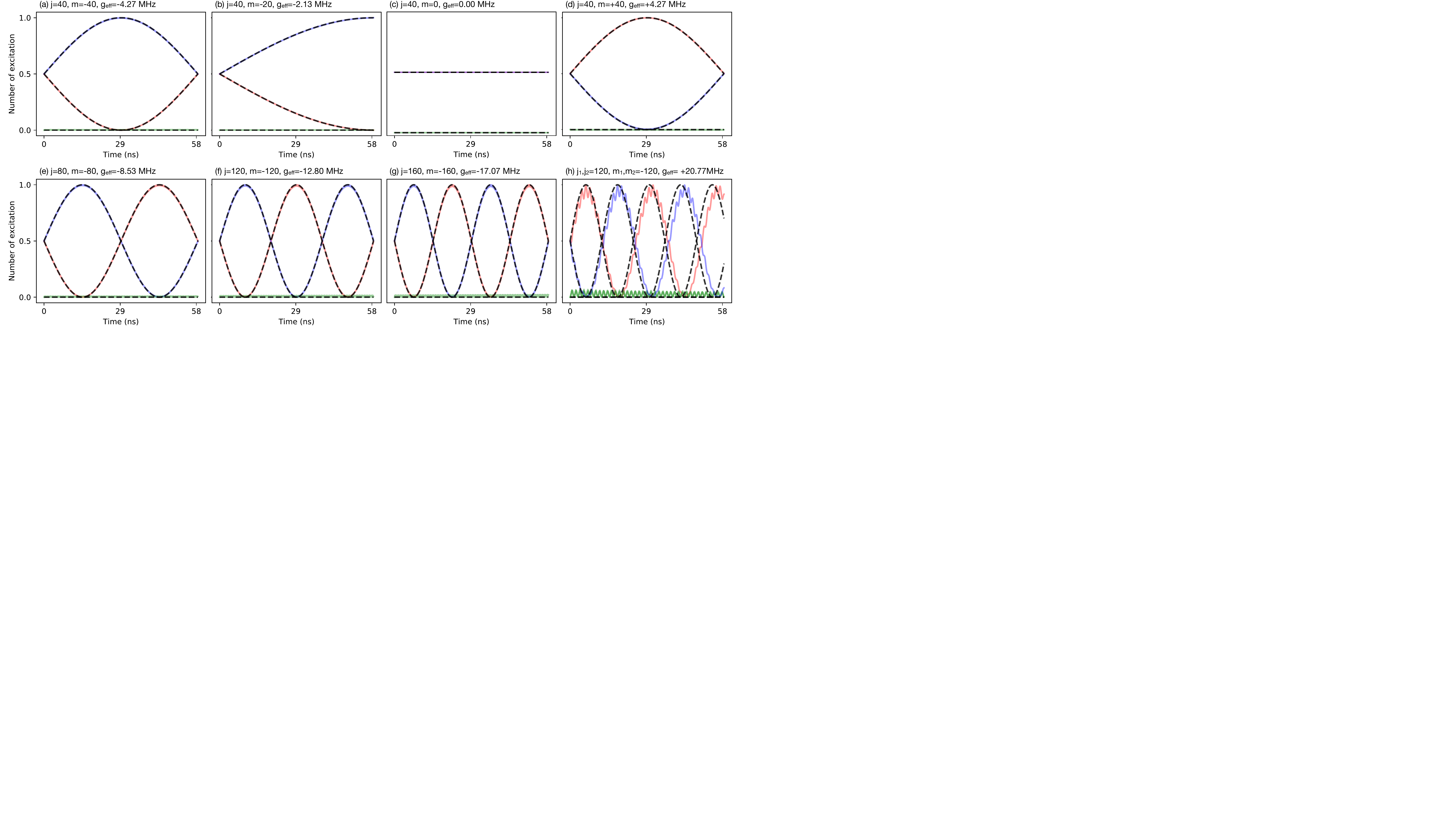}
  \caption{(a)-(d), When operating a $80$-qubit coupler with different magnetic quantum numbers, $m$, the two qubits under coupled exchange photons with different Rabi frequency. The interaction can be turned off when $m=0$, while the maximum coupling rate, $\pm 4.27\,{\rm MHz}$, is achieved when $m=\pm 40$. (e)-(g), Keeping the parameters unchanged while increasing the number of coupler qubits from $160$ to $320$, the effective coupling rate increases linearly from $-8.53\,{\rm MHz}$ through $-12.80\,{\rm MHz}$ to $-17.07\,{\rm MHz}$. In the last case, $\left|g_{\rm eff}\right|$ is $1.7$ times larger than the physical coupling rate between a single qubit and the coupler. (h), Dynamics of the system with a cascade of two couplers, where each layer contains $240$ homogeneous qubits. In all the panels, the red and blue curves correspond to the population of the two qubits under coupled, and green curves the magnetic quantum number of the qubit ensemble deviating from $m$. They are the simulation result with the original Hamiltonians, Eq.\,\eqref{eq:hamiltonian} and \eqref{eq:cascade}, while the dashed curves correspond to the predictions of the effective Hamiltonians.}
  \label{fig:simulation}
\end{figure*}

\section{Cascade of D-couplers}\label{sec:cascade}
The D-coupler may also be cascaded in a chain with several layers of qubit ensembles, as schematically shown in Fig.\,\ref{fig:schematic}(e). To simplify our discussion, we consider a system with $D$ layers of $N$ homogeneous qubits, where every two adjacent layers are coupled by an XY-type interaction
\begin{align}
	H &= \sum_{m=1}^{2} \omega_{m} x_m^{\dagger}x_m 
	+ \sum_{d=1}^{D}\frac{\omega_{\rm c}}{2}J_{d}^{z} \nonumber \\
	&+ g_{1}\left( x_{1}^{\dagger}+x_{1} \right)J_{1}^{x}
	+ g_{2}\left(x_{2}^{\dagger}+x_{2}\right)J_{D}^{x} \nonumber \\
	&+ \sum_{d=1}^{D-1}g_{\rm c} \left(J_{d}^{+}J_{d+1}^{-}+J_{d}^{-}J_{d+1}^{+}\right). \label{eq:cascade}
\end{align}
We focus on the case where all the qubits are initially prepared in the ground state, corresponding to the maximum effective coupling rate between $\rm X_1$ and $\rm X_2$. For large $N$, the collective qubits may be approximately described by a giant quantum oscillator with $J_d^{+} \approx \sqrt{N}a_d^{\dagger}$, $J_d^{-} \approx \sqrt{N}a_d$, and $J_d^{z} = 2a_d^{\dagger}a_d-N$ \cite{Katriel1986, Bullough1987, Bullough1989, Emary2003, Emary2003a}. 

Similar to the definition of magnons in a XY spin chain \cite{Karbach2005, Wojcik2005, Wojcik2007, Gratsea2018}, we diagonalize the coupler part of the Hamiltonian by introducing a collective operator 
\begin{align}
	a_k^{\pm} = \sqrt{\frac{2}{D+1}}\sum_{d=1}^{D}\sin\left(\frac{dk\pi}{D+1}\right)a_{d}^{\pm},\,\text{for}\,k=1,\cdots,D.
\end{align}
We are interested in the parameter regime where all the ``magnons" are largely off-resonant to the two circuits, i.e., $g_{m,k} \ll \Delta_{m,k}, \Sigma_{m,k}$ where $g_{m,k}=\sqrt{N}g_{m}\sin\left[mk\pi/(D+1)\right]\sqrt{2/(D+1)}$ is the coupling rate between $\rm X_m$ and the magnon-like mode, $a_k$, at frequency $\omega_k = \omega_{\rm c}+2g_{\rm c}N\cos\left[k\pi/(D+1)\right]$. Then, we apply a dispersive approximation (DA) to transform the component-coupler interactions into an effective interaction between $\rm X_1$ and $\rm X_2$. The effective Hamiltonian is too complicated to display here (see Appendix\,\ref{app:cascade} for detail), while the effective coupling rate is similar to the single-layer case 
\begin{align}
	g_{\rm eff} &= \frac{Ng_{1}g_{2}}{D+1}\sum_{m=1}^{2}\sum_{k=1}^{D}\sin\left(\frac{mk\pi}{D+1}\right)\left( \frac{1}{\Delta_{m,k}}-\frac{1}{\Sigma_{m,k}} \right). \label{eq:cascade_g}
\end{align}

Surprisingly, $g_{\rm eff}$ is almost independent of the number of layers, $D$. For a relatively small qubit number, $N$, in each layer, i.e., $Ng_{\rm c} \ll \Delta_{m,k}, \Sigma_{m,k}$, the value of $g_{\rm eff}$ scales quadratically with $N$. With the increase of $N$, the exponent of the power law increases monotonically until DA breaks down. In this case, one may enlarge the detuning frequency and add more layers and qubits at the same time to achieve a balance between the amplification rate and the validity of the effective model. However, we should note that an Ising-type interaction between two adjacent layers causes a correction in Eq.\,\eqref{eq:cascade_g}, as revealed in Appendix\,\ref{app:cascade}, which may lead to a larger $g_{\rm eff}$ in certain parameter regimes. Besides, a varying-frequency design of the coupler qubits among different layers may also lead to a different expression of $g_{\rm eff}$.

\section{Simulation results}
For illustration, we simulate the dynamics of a system which consists of two qubits and a D-coupler with $80$ qubits, as shown in Fig.\,\ref{fig:simulation}(a)-(d). The two qubits are resonant at $1\,{\rm GHz}$ and largely detuned from the coupler qubits that are fixed at $4\,{\rm GHz}$. The physical coupling rate, $g_{m}$, is set as $+10\,{\rm MHz}$, which fulfills the requirements of both DA and RWA. The possibly small qubit crosstalk inside the coupler is neglected for simplisity. From (a) to (d), we control the effective coupling rate, from $-4.27\,{\rm MHz}$ to $+4.27\,{\rm MHz}$, by engineering the qubit ensemble in different collective states. The coupling is completely turned off in (c), when the magnetic quantum number is \textit{zero}. It also changes the sign comparing (a) and (d), which corresponds to the different signs of the magnetic quantum numbers. 

Next, we keep all the coupler qubits in the ground state and increase the qubit number to $160$, $240$, and $320$, as shown in Fig.\,\ref{fig:simulation}(e)-(g). Correspondingly, the effective coupling rate increases linearly from $-8.53$ through $-12.80$ to $-17.07\,{\rm MHz}$. Here, the maximum value of $g_{\rm eff}$ is $1.7$ times larger than the physical coupling rate, $g_{m}$, while the effective Hamiltonian, Eq.\,\eqref{eq:effective}, still faithfully describes the dynamics of the system. This observation clearly demonstrates that an amplification of the interactions can be achieved in superconducting quantum circuits, which has not yet been reported in the literature. However, one may also note a small and periodic excitation of the coupler for increasingly large $N$, which is mainly caused by the finite ratio of $g_{m}/\Delta_{m}$ that limits the accuracy of the DA. These ripples can be effectively suppressed by increasing the detuning frequency and the qubit number by the same scaling factor, while keeping the effective coupling rate unchanged. 

In Fig.\,\ref{fig:simulation}(h), we simulate the system dynamics with a two-layer coupler, where each layer contains $240$ collective qubits. The coupling rate between two adjacent layers, $g_{\rm q}$, is assumed to be $+10\,{\rm MHz}$, while the other parameters are set identical to (a)-(g). Regardless of the noticeable ripples in the excitation numbers, we obtain an effective coupling rate of around $+20.77\,{\rm MHz}$, which is much larger than that for the single layer case shown in (f). The sign of the coupling rate is also different from (f). However, considering the total number of $480$ qubits in the two-layer coupler, adding more layers may not amplify the interactions as efficiently as increasing the number of qubits in a single layer, which is estimated to be $-25.60\,{\rm MHz}$ for the same number of qubits. Moreover, the amplification rate may even decrease when adding an addition layer in certain scenarios. For example, we obtain an effective coupling rate of $+5.12\,{\rm MHz}$ for two layers of $160$ qubits, while it is $-8.53\,{\rm MHz}$ for a single layer of $160$ qubits, as shown in (e).

\section{Conclusions $\rm \&$ outlook}
We have proposed a so-called D-coupler based on the Dicke model, and showed that it can be used to tune and amplify the effective interactions between two general circuits. The effective coupling rate is controlled by the collective magnetic quantum number of the qubit ensemble, and it is free from collective decay and decoherence when the qubits are prepared in the subradiant states. The tuning procedure, from any initial to final values, is achieved in a single step by applying single-photon $\pi$-pulses, without going through any intermediate values. The dynamic range scales linearly with the number of qubits, where the maximum can be made larger than the physical coupling rates between the coupler and the circuits.

We have also discussed a multi-layer design of the D-coupler. The circuit configuration is reminiscent to a traveling wave parametric amplifier (TWPA), which may be seen as a \textit{series} configuration of the coupler qubits. In this perspective, amplification from the input to output fields may happen in the presence of a parametric driving field, when the effective coupling rate between the end qubits exceeds a threshold relating to their external decay rates \cite{Clerk2010}. However, our result may indicate that a \textit{parallel} configuration of the coupler qubits may be an alternative and more efficient way to achieve quantum limited amplification at the same number of junctions. 

Despite the collective properties, one major difference between the D-coupler and most of the existing proposals is that the effective coupling rate is controlled by the quantum state of the coupler but not the detuning frequency. Hence, the coupler can be made of only one or two fixed-frequency qubits, which significantly simplifies the sample fabrication, and, most importantly, leads to a flux-free design of superconducting quantum circuits where the qubit properties can be optimized with regard to only one noise source, i.e., the charge noise. One may also consider engineering the coupler state in the steady states of a driven-dissipative setup, which relaxes the assumption of a collective reservoir and may result into a more robust implementation. An experimental realization of our protocol could offer significant advantages in designing superconducting quantum circuits for quantum information processing.

\section*{Acknowledgements}
We thank David Castells-Graells and Cosimo C. Rusconi for fruitful discussions. We use the QUANTUM add-on for Mathematica \cite{Munoz2016, Mathematica} for deriving the effective Hamiltonian, and the QuTiP package in Python \cite{Johansson2013} for simulation. We acknowledge support by the German Research Foundation via Germany's Excellence Strategy (No.\,EXC-2111-390814868), the Elite Network of Bavaria through the program ExQM, and the European Union via the Quantum Flagship project QMiCS (No.\,820505).

\appendix
\begin{widetext}
\section{Effective Hamiltonian with one layer}\label{app:one}
We consider a system, $H=H_0+V$, where two general circuit components are coupled via a qubit ensemble
\begin{align}
	H_0 &= \sum_{m=1}^{2}\omega_{m} x_m^{\dagger}x_m 
	+ \sum_{n=1}^{N}\frac{\omega_{n}}{2}\sigma_{n}^{z}
	+ \sum_{\{n, n'\}} g_{n,n'} \left(\sigma_{n}^{+}\sigma_{n'}^{-} + \sigma_{n}^{-}\sigma_{n'}^{+}\right), \\
	V &= \sum_{m=1}^{2}\sum_{n=1}^{N}\underbrace{g_{m,n}\left( x_{m}^{\dagger}\sigma_{n}^{-} + x_{m}\sigma_{n}^{+} \right)}_{V_1}
	+ \underbrace{g_{m,n}\left(x_{m}^{\dagger}\sigma_{n}^{+}+x_{m}\sigma_{n}^{-}\right)}_{V_2}.
\end{align}
Here, $\omega_{m}$, $x_m$, and $x_m^{\dagger}$ are the resonant frequency, raising and lowering operators of the $m$th system component $\rm X_m$. Furthermore, $\omega_{n}$ and $\sigma_{n}^{\alpha}$ with $\alpha=x,y,z$ are the characteristic frequency and standard Pauli operators of the $n$th qubit in the ensemble with a total qubit number of $N$, $g_{\alpha,\beta}$ is the coupling rate between the two components $\alpha$ and $\beta$, and $\{n, n'\}$ takes all possible pairs in the ensemble. To transform the component-coupler interaction into an effective component-component interaction to the second order of $g_{\alpha,\beta}$, we apply the following unitary transformation to the original Hamiltonian \cite{Zueco2009}
\begin{equation}
	U = \exp\Bigg[\overbrace{\sum_{m=1}^{2}\sum_{n=1}^{N}-\frac{g_{m,n}}{\Delta_{m,n}}\left( x_m^{\dagger}\sigma_{n}^{-} - x_m \sigma_{n}^{+} \right)}^{X_1}
	+ \overbrace{\sum_{m=1}^{2}\sum_{n=1}^{N}-\frac{g_{m,n}}{\Sigma_{m,n}}\left( x_m^{\dagger}\sigma_{n}^{+} - x_m \sigma_{n}^{-} \right)}^{X_2}\Bigg].
\end{equation}
On the one hand, we have $\llbracket H_0,X_1+X_2\rrbracket=-V$. The transformation can be simplified as $U^{\dagger}HU=H_0 + (1/2)\llbracket V, X_1+X_2 \rrbracket$ to the second-order accuracy of $g_{m,n}/\Delta_{m,n}$ and $g_{m,n}/\Sigma_{m,n}$. On the other hand, we have
\begin{align}
	\llbracket V_1, X_1\rrbracket &= -\frac{g_{m,n}}{\Delta_{m,n}}\left\{ g_{m',n}\sigma_{n}^{z}\left( x_1^{\dagger}x_2 + x_1^{\dagger}x_2 \right)
	+ g_{m,n}\left(\sigma_{n}^{z}\left\{x_m,x_m^{\dagger} \right\} + \left[x_m,x_m^{\dagger}\right] \right)
	+ g_{m,n'}\left[x_m,x_m^{\dagger}\right]\left( \sigma_{n}^{+}\sigma_{n'}^{-} + \sigma_{n}^{-}\sigma_{n'}^{+} \right)\right\}, \\
	\llbracket V_1, X_2\rrbracket &= \frac{g_{m,n}}{\Sigma_{m,n}}\left\{g_{m',n}\sigma_{n}^{z}\left( x_1^{\dagger}x_2^{\dagger} + x_1x_2 \right)
	+ g_{m,n}\sigma_{n}^{z} \left(x_m^{\dagger 2} + x_m^2 \right)
	- g_{m,n'}\left[x_m,x_m^{\dagger}\right]\left( \sigma_{n}^{+}\sigma_{n'}^{+} + \sigma_{n}^{-}\sigma_{n'}^{-} \right)\right\}, \\
	\llbracket V_2, X_1\rrbracket &= 
	-\frac{g_{m,n}}{\Delta_{m,n}}\left\{g_{m',n}\sigma_{n}^{z}\left( x_1^{\dagger}x_2^{\dagger} + x_1x_2 \right)
	+ g_{m,n}\sigma_{n}^{z} \left(x_m^{\dagger 2} + x_m^2 \right)
	+ g_{m,n'}\left[x_m,x_m^{\dagger}\right]\left( \sigma_{n}^{+}\sigma_{n'}^{+} + \sigma_{n}^{-}\sigma_{n'}^{-} \right)\right\}, \\
	\llbracket V_2, X_2\rrbracket &= \frac{g_{m,n}}{\Sigma_{m,n}}\left\{g_{m',n}\sigma_{n}^{z}\left( x_1^{\dagger}x_2 + x_1^{\dagger}x_2 \right)
	+ g_{m,n}\left(\sigma_{n}^{z}\left\{x_m,x_m^{\dagger} \right\} - \left[x_m,x_m^{\dagger}\right] \right)
	- g_{m,n'}\left[x_m,x_m^{\dagger}\right]\left( \sigma_{n}^{+}\sigma_{n'}^{-} + \sigma_{n}^{-}\sigma_{n'}^{+} \right)\right\}.
\end{align}
Here, we have omitted the summation symbols over $m$ and $n$ for the simplicity of notation, $m'$ and $n'$ indicate all the different numbers from $m$ and $n$. In total, we obtain the effective Hamiltonian
\begin{align}
	\tilde{H} &= \omega_{m} x_m^{\dagger}x_m 
	+ \frac{\omega_{n}}{2}\sigma_{n}^{z}
	+ \frac{g_{n,n'}}{2} \sigma_{n}^{x}\sigma_{n'}^{x} 
	+ \frac{g_{m,n}g_{m',n}}{2}\left(\frac{1}{\Sigma_{m,n}}- \frac{1}{\Delta_{m,n}}\right)
	\sigma_{n}^{z}\left( r_1^{\dagger} + r_1 \right)\left( r_2^{\dagger} + r_2 \right) \nonumber \\
	&+ \frac{g_{m,n}^2}{2}\left(\frac{1}{\Sigma_{m,n}}-\frac{1}{\Delta_{m,n}}\right)\sigma_{n}^{z}\left(x_1 + x_1^{\dagger}\right)^2 
	- \frac{g_{m,n}^2}{2}\left(\frac{1}{\Sigma_{m,n}}+\frac{1}{\Delta_{m,n}}\right)\llbracket x_m, x_m^{\dagger}\rrbracket \left(\sigma_n^{-} + \sigma_n^{+}\right)^2 \nonumber \\
	&- \frac{g_{m,n}g_{m,n'}}{2}\left(\frac{1}{\Sigma_{m,n}}+\frac{1}{\Delta_{m,n}}\right)\llbracket x_m, x_m^{\dagger}\rrbracket \sigma_{n}^{x}\sigma_{n'}^{x}.
\end{align}

For the cases with two resonators, one resonator and one qubit, and two qubits, respectively, we have the following effective Hamiltonians
{\small\begin{align}
	\tilde{H}_{\text{R-R}} &= \sum_{m=1}^{2}\sum_{n=1}^{N} \left[\omega_{m} + g_{m,n}^2\left( \frac{1}{\Sigma_{m,n}}-\frac{1}{\Delta_{m,n}}\right)\sigma_{n}^{z} \right]r_m^{\dagger}r_m 
	+ \frac{g_{1,n}g_{2,n}}{2}\left(\frac{1}{\Sigma_{m,n}}- \frac{1}{\Delta_{m,n}}\right)\sigma_{n}^{z}\left( r_1^{\dagger} + r_1 \right)\left( r_2^{\dagger} + r_2 \right) \nonumber \\
	&+ \frac{1}{2}\left[\omega_{n} 
	+ g_{m,n}^2\left( \frac{1}{\Sigma_{m,n}}-\frac{1}{\Delta_{m,n}}\right) \left(r_m^{\dagger}r_m^{\dagger}+r_m r_m + 1\right)\right] \sigma_{n}^{z} \nonumber \\
	&+ \sum_{n=1}^{N}\sum_{n'\neq n}\frac{1}{2}\left[g_{n,n'}-g_{m,n}g_{m,n'}\left(\frac{1}{\Sigma_{m,n}}+\frac{1}{\Delta_{m,n}}\right)\right]\sigma_{n}^{x}\sigma_{n'}^{x}, \\
	\tilde{H}_{\text{R-Q}} &= \sum_{m=1}^{2}\sum_{n=1}^{N} \left[\omega_{1} + g_{1,n}^2\left( \frac{1}{\Sigma_{1,n}}-\frac{1}{\Delta_{1,n}}\right)\sigma_{n}^{z} \right]r_1^{\dagger}r_1
	+ \left[\omega_{2}+g_{2,n}^2\left(\frac{1}{\Sigma_{2,n}}+\frac{1}{\Delta_{2,n}}\right)\right]\frac{\sigma_{2}^{z}}{2} \nonumber \\
	&+ \frac{g_{1,n}g_{2,n}}{2}\left(\frac{1}{\Sigma_{m,n}}- \frac{1}{\Delta_{m,n}}\right)\sigma_{n}^{z}\left( r_1^{\dagger} + r_1\right) \sigma_{2}^{x}\nonumber \\
	&+ \frac{1}{2}\left[ \omega_{n} + g_{1,n}^2\left( \frac{1}{\Sigma_{1,n}}-\frac{1}{\Delta_{1,n}}\right) \left(r_1^{\dagger}r_1^{\dagger}+r_1 r_1 + 1\right) 
	+ g_{2,n}^2\left(\frac{1}{\Sigma_{2,n}} - \frac{1}{\Delta_{2,n}}\right) \right]\sigma_{n}^{z} \nonumber \\
	&+ \sum_{n=1}^{N}\sum_{n'\neq n}\frac{1}{2}\left[ g_{n,n'} 
	- g_{1,n}g_{1,n'}\left(\frac{1}{\Sigma_{1,n}}+ \frac{1}{\Delta_{1,n}}\right)
	+ g_{2,n}g_{2,n'}\left(\frac{1}{\Sigma_{2,n}}+ \frac{1}{\Delta_{2,n}}\right)\sigma_{2}^{z} \right]\sigma_{n}^{x}\sigma_{n'}^{x}, \\
	\tilde{H}_{\text{Q-Q}} &= \sum_{m=1}^{2}\sum_{n=1}^{N} \left[\omega_{m} + g_{m,n}^2\left(\frac{1}{\Sigma_{m,n}} + \frac{1}{\Delta_{m,n}}\right)\right]\frac{\sigma_{m}^{z}}{2} 
	+ \frac{g_{1,n}g_{2,n}}{2}\left(\frac{1}{\Sigma_{m,n}}- \frac{1}{\Delta_{m,n}}\right)\sigma_{n}^{z}\sigma_{1}^{x}\sigma_{2}^{x} \nonumber \\
	&+ \frac{1}{2}\left[\omega_{n}+g_{m,n}^2\left(\frac{1}{\Sigma_{m,n}}-\frac{1}{\Delta_{m,n}}\right)\right]\sigma_n^{z}
	+ \sum_{n=1}^{N}\sum_{n'\neq n}\frac{1}{2}\left[g_{n,n'} + g_{m,n}g_{m,n'}\left(\frac{1}{\Sigma_{m,n}}+\frac{1}{\Delta_{m,n}}\right)\sigma_{m}^{z}\right]\sigma_{n}^{x}\sigma_{n'}^{x}. 
\end{align}}%
If we further assume the homogeneity among the coupler qubits, we use the  collective angular momentum operators to describe the whole qubit ensemble and arrive at a more compact form of the effective Hamiltonian. They are
{\small\begin{align}
	\tilde{H}_{\text{R-R}} &= \sum_{m=1}^{2}\left[\omega_{m} + g_{m}^2\left( \frac{1}{\Sigma_{m}}-\frac{1}{\Delta_{m}}\right)J^{z} \right]r_m^{\dagger}r_m 
	+ \frac{g_1g_2}{2}\left(\frac{1}{\Sigma_{m}}- \frac{1}{\Delta_{m}}\right)J^{z}\left( r_1^{\dagger} + r_1 \right)\left( r_2^{\dagger} + r_2 \right) \nonumber \\
	&+ \frac{1}{2}\left[\omega_{\rm c} 
	+ g_{m}^2\left( \frac{1}{\Sigma_{m}}-\frac{1}{\Delta_{m}}\right) \left(r_m^{\dagger}r_m^{\dagger}+r_m r_m + 1\right)\right]J^{z} 
	+\frac{1}{2}\left[g_{\rm c}-g_{m}^2\left(\frac{1}{\Sigma_{m}}+\frac{1}{\Delta_{m}}\right)\right]\left(J^{x}\right)^{2}, \\
	\tilde{H}_{\text{R-Q}} &= \sum_{m=1}^{2}\left[\omega_{1} + g_{1}^2\left( \frac{1}{\Sigma_{1}}-\frac{1}{\Delta_{1}}\right)J^{z} \right]r_1^{\dagger}r_1
	+ \left[\omega_{2}+g_{2}^2\left(\frac{1}{\Sigma_{2}}+\frac{1}{\Delta_{2}}\right)\left(J^{x}\right)^2\right]\frac{\sigma_{2}^{z}}{2} 
	+\frac{g_1g_2}{2}\left(\frac{1}{\Sigma_{m}}- \frac{1}{\Delta_{m}}\right)J^{z}\left( r_1^{\dagger} + r_1\right) \sigma_{2}^{x}\nonumber \\
	&+ \frac{1}{2}\left[\omega_{\rm c} + g_{1}^2\left( \frac{1}{\Sigma_{1}}-\frac{1}{\Delta_{1}}\right) \left(r_1^{\dagger}r_1^{\dagger}+r_1 r_1 + 1\right) 
	+ g_{2}^2\left(\frac{1}{\Sigma_{2}} - \frac{1}{\Delta_{2}}\right) \right]J^{z} \nonumber \\
	&+\frac{1}{2}\left[ g_{\rm c} 
	- g_{1}^2\left(\frac{1}{\Sigma_{1}}+ \frac{1}{\Delta_{1}}\right) \right]\left(J^{x}\right)^{2}, \\
	\tilde{H}_{\text{Q-Q}} &= \sum_{m=1}^{2}\left[\omega_{m} + g_{m}^2\left(\frac{1}{\Sigma_{m}} + \frac{1}{\Delta_{m}}\right)\left(J^{x}\right)^2\right]\frac{\sigma_{m}^{z}}{2} 
	+\frac{g_1g_2}{2}\left(\frac{1}{\Sigma_{m}}- \frac{1}{\Delta_{m}}\right) J^{z}\sigma_{1}^{x}\sigma_{2}^{x} \nonumber \\
	&+\frac{1}{2}\left[\omega_{\rm c}+g_{m}^2\left(\frac{1}{\Sigma_{m}}-\frac{1}{\Delta_{m}}\right)\right]J^{z}
	+\frac{1}{2}g_{\rm c} \left(J^{x}\right)^{2}. 
\end{align}}%

\section{Effective Hamiltonian with multiple layers}\label{app:cascade}
We consider a system with $D$ layers of $N$ homogeneous qubits, where any two adjacent layers are coupled by an XY-type interaction
\begin{align}
	H &= \sum_{m=1}^{2} \omega_{m} x_m^{\dagger}x_m 
	+ \sum_{d=1}^{D}\frac{\omega_{\rm c}}{2}J_{d}^{z}
	+ \sum_{d=1}^{D-1}g_{\rm c} \left(J_{d}^{+}J_{d+1}^{-}+J_{d}^{-}J_{d+1}^{+}\right)
	+ g_{1}\left( x_{1}^{\dagger}+x_{1} \right)J_{1}^{x}
	+ g_{2}\left(x_{2}^{\dagger}+x_{2}\right)J_{d}^{x}.
\end{align}
By assuming that $\langle J_{d}^{z} \rangle \approx -N/2$, we introduce the following replacement of variables for large $N$ \cite{Katriel1986, Bullough1987, Bullough1989, Emary2003, Emary2003a}
\begin{align}
	J_{d}^{+} \approx \sqrt{N}a_{d}^{\dagger},\,
	J_{d}^{-} \approx \sqrt{N}a_{d},\,
	J_{d}^{z} = 2a_{d}^{\dagger}a_{d}-N.
\end{align}
This gives
\begin{align}
	H &= \sum_{m=1}^{2}\omega_{m} x_m^{\dagger}x_m 
	+ \sqrt{N}g_{1}\left( x_{1}^{\dagger} + x_{1} \right)\left( a_{1}^{\dagger} + a_{1} \right)
	+ \sqrt{N}g_{2}\left( x_{2}^{\dagger}+ x_{2} \right)\left( a_{d}^{\dagger} + a_{d} \right) \nonumber \\
	&+ \sum_{d=1}^{D} \omega_{\rm c} a_{d}^{\dagger}a_{d}
	+ \sum_{d=1}^{D-1} Ng_{\rm q} \left(a_{d}^{\dagger}a_{d+1}+a_{d}a_{d+1}^{\dagger}\right).
\end{align}
We note that the last term should be written as $\sum_{d=1}^{D-1} Ng_{\rm q} \left(a_{d}^{\dagger}+a_{d}\right)\left(a_{d+1}^{\dagger}+ a_{d+1}\right)$ for an Ising-type interaction, described by $g_{\rm c}J_{d}^{x}J_{d+1}^{x}$, between two adjacent layers.

Similar to the definition of magnons in a XY-type spin chain \cite{Karbach2005, Wojcik2005, Wojcik2007, Gratsea2018}, we define 
\begin{align}
	a_{k}^{\pm} = \sqrt{\frac{2}{D+1}}\sum_{d=1}^{D}\sin\left(\frac{sk\pi}{D+1}\right)a_{d}^{\pm}.
\end{align}
The Hamiltonian can be written as
\begin{align}
	H_{0} &= \sum_{m=1}^{2}\omega_{m} x_m^{\dagger}x_m 
	+ \sum_{k=1}^{D}\left(\omega_{\rm c}+2g_k\right)a_{k}^{\dagger}a_{k}, \\
	V &= \sum_{m=1}^{2}\sum_{k=1}^{N} \underbrace{g_{m,k}\left( x_{m}^{\dagger}a_{k} + x_{m} a_{k}^{\dagger} \right)}_{V_1} 
	+ \underbrace{g_{m,k}\left( x_{m}^{\dagger}a_{k}^{\dagger} + x_{m} a_{k} \right)}_{V_2}, 
\end{align}
where $g_{k}=Ng_{\rm q}\cos\left[k\pi/(D+1)\right]$, $g_{m,k}=\sqrt{N}g_{m}\sin\left[mk\pi/(D+1)\right]\sqrt{2/(D+1)}$. For an Ising-type interaction, one may add $\sum_{k=1}^{D} g_{k} \left(a_{k}^{\dagger 2} + a_{k}^{2}\right)$ in $H_0$.

To derive the effective Hamiltonian, we apply the following unitary transformation
\begin{align}
	U &= \exp\Bigg[\overbrace{\sum_{m=1}^{2}\sum_{k=1}^{D}\lambda_{m,k}^{-}\left( x_m^{\dagger}a_{k} - x_m a_{k}^{\dagger} \right)}^{X_1}
	+\overbrace{\sum_{m=1}^{2}\sum_{k=1}^{D}\lambda_{m,k}^{+}\left( x_m^{\dagger}a_{k}^{\dagger} - x_m a_{k} \right)}^{X_2}\Bigg].
\end{align}
We obtain
{\small\begin{align}
	\llbracket H_0, X_1+X_2 \rrbracket 
	&= \lambda_{m,k}^{-}\left(\Delta_{m,n} - 2g_k\right) 
	\left( x_m^{+}a_{k} + x_m^{-}a_{k}^{\dagger} \right) 
	+ \lambda_{m,k}^{+}\left(\Sigma_{m,n} + 2 g_k\right) 
	\left( x_m^{+}a_{k}^{\dagger} + x_m^{-}a_{k} \right), \\
	\llbracket H_0, X_1+X_2 \rrbracket 
	&= \left[\lambda_{m,k}^{-}\Delta_{m,n} 
	+ 2\left(\lambda_{m,k}^{+}-\lambda_{m,k}^{-}\right) g_k \right] 
	\left( x_m^{+}a_{k} + x_m^{-}a_{k}^{\dagger} \right) 
	+ \left[\lambda_{m,k}^{+}\Sigma_{m,n} 
	+ 2\left(\lambda_{m,k}^{+}-\lambda_{m,k}^{-}\right) g_k\right] 
	\left( x_m^{+}a_{k}^{\dagger} + x_m^{-}a_{k} \right),
\end{align}%
for the XY- and Ising-type interactions, respectively. The rest of the commutators are
\begin{align}
	\llbracket V_1, X_1 \rrbracket 
	&= -\lambda_{m,k}^{-}\left\{g_{m',k}
	\left( x_1^{\dagger}x_2 + x_1x_2^{\dagger} \right)
	+ g_{m,k} \left\{x_m,x_m^{\dagger}\right\} 
	- \left(g_{m,k}\left\{a_{k}^{\dagger},a_{k} \right\} + g_{m,k'}\left\{ a_{k}^{\dagger}a_{k'} + a_{k}a_{k'}^{\dagger} \right\}\right) \llbracket x_m,x_m^{\dagger}\rrbracket \right\}\\
	\llbracket V_1, X_2 \rrbracket 
	&= \lambda_{m,k}^{+}\left\{g_{m',k}
	\left( x_1x_2 + x_1^{\dagger}x_2^{\dagger} \right)	
	+ g_{m,k} \left\{x_m^2 + x_m^{\dagger 2}\right\} 
	+ \left(g_{m,k}\left\{a_{k}^{+ 2}+a_{k}^{- 2} \right\} + g_{m,k'}\left\{ a_{k}^{\dagger}a_{k'} + a_{k}a_{k'}^{\dagger} \right\}\right) \llbracket x_m,x_m^{\dagger}\rrbracket \right\},\\
	\llbracket V_2, X_1 \rrbracket
	&= -\lambda_{m,k}^{-} \left\{g_{m',k}
	\left( x_1x_2 + x_1^{\dagger}x_2^{\dagger} \right)
	+ g_{m,k} \left\{x_m^2 + x_m^{\dagger 2}\right\} 
	- \left(g_{m,k}\left\{a_{k}^{+ 2}+a_{k}^{- 2} \right\} + g_{m,k'}\left\{ a_{k}^{\dagger}a_{k'} + a_{k}a_{k'}^{\dagger} \right\}\right) \llbracket x_m,x_m^{\dagger}\rrbracket \right\}, \\
	\llbracket V_2, X_2 \rrbracket 
	&= \lambda_{m,k}^{+}\left\{ g_{m',k}
	\left( x_1^{\dagger}x_2 + x_1x_2^{\dagger} \right)
	+ g_{m,k} \left\{x_m,x_m^{\dagger}\right\} 
	+ \left(g_{m,k}\left\{a_{k}^{\dagger},a_{k} \right\} + g_{m,k'}\left\{ a_{k}^{\dagger}a_{k'} + a_{k}a_{k'}^{\dagger} \right\}\right) \llbracket x_m,x_m^{\dagger}\rrbracket \right\}.
\end{align}}%
As before, we have omitted the summation symbol in the above formulae for the simplicity of notation. The component-coupler interaction can be eliminated to the second-order accuracy if 
\begin{align}
	&\lambda_{m,k}^{-} = -\frac{g_{m,n}}{\left(\Delta_{m,n} - 2g_k \right)},\,
	\lambda_{m,k}^{+} = -\frac{g_{m,n}}{\left(\Sigma_{m,n} + 2g_k \right)},
\end{align}
or 
\begin{align}
	&\lambda_{m,k}^{-}\Delta_{m,n} 
	+ 2\left(\lambda_{m,k}^{+}-\lambda_{m,k}^{-}\right) g_k = -g_{m,n},\,
	\lambda_{m,k}^{+}\Sigma_{m,n} 
	+ 2\left(\lambda_{m,k}^{+}-\lambda_{m,k}^{-}\right) g_k = -g_{m,n},
\end{align}
which gives the following effective Hamiltonian
\begin{align}
	\tilde{H} &= \omega_{m} x_m^{\dagger}x_m 
	+ \left(\omega_{\rm c}+2g_k\right) a_{k}^{\dagger} a_{k}
	+ \underline{g_k \left(a_{k}^{\dagger 2} + a_{k}^2\right)} \nonumber \\
	&+ \frac{1}{2}\left(\lambda_{m,k}^{+}-\lambda_{m,k}^{-}\right)g_{m',k}
	\left(x_1+x_1^{\dagger}\right)\left(x_2+x_2^{\dagger}\right)	
	+ \frac{1}{2}\left(\lambda_{m,k}^{+}-\lambda_{m,k}^{-}\right)g_{m,k}\left(x_m + x_m^{\dagger}\right)^2 \nonumber \\
	&+ \frac{1}{2}\left(\lambda_{m,k}^{+}+\lambda_{m,k}^{-}\right)g_{m,k}\left(a_{k}+a_{k}^{\dagger}\right)^2 \llbracket x_m,x_m^{\dagger}\rrbracket
	+ \frac{1}{2}\left(\lambda_{m,k}^{+}+\lambda_{m,k}^{-}\right)g_{m,k'}\left(a_{k}^{\dagger} + a_{k} \right)\left(a_{k'}^{\dagger} + a_{k'} \right)\llbracket x_m,x_m^{\dagger}\rrbracket.
\end{align}
Here, the underlined term exists only for an Ising-type interaction.
\end{widetext}

\bibliography{DCoupler_REF}  
\end{document}